\documentclass[a4paper,12pt]{article}
\usepackage{amsmath,amssymb}
\usepackage{graphicx}

\DeclareMathOperator{\sgn}{sgn}
\DeclareMathOperator{\Tr}{Tr}

\numberwithin{equation}{section}
\allowdisplaybreaks

\usepackage{geometry}
\newcounter{aff}

\begin{document}

\begin{titlepage}
%% change the footnote symbol
\renewcommand{\thefootnote}{\fnsymbol{footnote}}

\begin{flushright}
{\footnotesize YITP-15-106}
\end{flushright}
\begin{center}
{\Large\bf Large $N$ behavior of mass deformed ABJM theory
\\[6pt]}

\bigskip\bigskip
{\large 
Tomoki Nosaka,\footnote[1]{\tt nosaka(at)yukawa.kyoto-u.ac.jp}
\quad
Kazuma Shimizu\footnote[2]{\tt kazuma.shimizu(at)yukawa.kyoto-u.ac.jp}
\quad and \quad
Seiji Terashima\footnote[3]{\tt terasima(at)yukawa.kyoto-u.ac.jp}
}\\
\bigskip
{\small\it Yukawa Institute for Theoretical Physics,
Kyoto University,\\
Kyoto 606-8502, Japan}
\end{center}

\begin{abstract}

In this paper, using the localization technique we analyze the large $N$ limit of the mass deformed Aharony-Bergman-Jafferis-Maldacena (ABJM) theory on the three sphere with a finite mass parameter and finite Chern-Simons levels.
We find two different solutions of the saddle point equations in the large $N$ limit.
With these solutions we compute the free energy and find that there is a first order phase transition.
Our results may predict a phase transition in the dual gravity theory.

\end{abstract}

\bigskip\bigskip\bigskip

\centering

\end{titlepage}
\tableofcontents

\section{Introduction}
To understand the M-theory, it should be important to study the M2-branes and the M5-branes.
Recently, it was proposed that the stack of $N$ M2-branes on the orbifold $\mathbb{C}^4/\mathbb{Z}_k$ $(k=1,2,\cdots )$ can be studied by a three dimensional $\text{U}(N)\times \text{U}(N)$ superconformal Chern-Simons matter theory \cite{ABJM}, which is called the ABJM theory.
Similarly, the M5-branes are believed to be described by some non-trivial six dimensional field theory.
The field theory on the M5-branes itself would also be interesting and play important roles to understand various non-perturbative properties of supersymmetric gauge theories.
However, there are no explicit field theoretical descriptions for the multiple M5-branes so far.

On the other hand, the M5-branes can be realized as the solitons in the ABJM theory.
Indeed we can construct BPS solitons which are the analogues of the Nahm data for the non-abelian monopoles in $(3+1)$ dimensional gauge theory \cite{T,NT,ST}.\footnote{
This structure is an extension of the work \cite{BH} which motivate the early proposal of the field theory on multiple M2-branes \cite{BL1,G,BL2,BL3} (see also \cite{GRVV,HL}).
}
In these solutions the M5-branes are realized as the M2-branes blowing up into a fuzzy sphere.
There is also a noncommutative-plane-like construction of the M5-branes \cite{TY1,TY2}.
These solutions enable us to investigate the M5-branes via the ABJM theory.

Similar fuzzy sphere structure appears also as non-trivial vacua in the mass deformed ABJM theory \cite{GRVV} (see also \cite{NPR}).
This configuration does not carry net M5-brane charges, but carries the dipole M5-brane charges.
This is an analogue of the Myers effect for the D-branes \cite{M}, which is, of course, related to the Nahm construction of the monopoles.
The mass deformed ABJM theory should also be a useful tool to study the M5-branes.

Relation to the M5-branes is also observed in the gravity side.
The UV limit of the mass deformed ABJM theory is same as the ABJM theory itself.
On the other hand, the mass deformation breaks the conformal invariance and hence should result in the different IR behavior.
Indeed the analyses of the holographic RG flow \cite{BW,PTW} suggest that the theory would describe a particular configuration of the M5-branes in the IR limit.
Hence it is interesting to study the mass deformed ABJM theory for large $N$.

In this paper, as a simple example we shall study the partition function (free energy) of the mass deformed ABJM theory on $S^3$ in the large $N$ limit.
With the help of the localization technique \cite{P,W,N}, the partition function can be exactly computed by a $2N$ dimensional matrix model \cite{KWY2,J,HHL}.
Though it is still difficult to perform these integrals for general large number $N$, in the large $N$ limit we can evaluate the partition function by the saddle point approximation.
We achieve to solve the saddle point equations for finite values of the Chern-Simons levels and the mass deformation parameter, i.e. in the M-theoretical regime.
With the solutions (eigenvalue distributions), we compute the exact large $N$ partition function.

Interestingly, we find two different solutions, which cause a first order phase transition in the large $N$ limit.
Since the theory we consider coincide to the ABJM theory in the UV limit, the dual geometry will be characterized by the same asymptotics as in the ABJM case: it will be asymptotically $\text{AdS}_4\times S^7/\mathbb{Z}_k$ where the boundary of $\mathrm{AdS}_4$ is $S^3$.
Our result may predict the existence of new solutions of this class and a phase transition in the gravity side.

Among various mass deformations of the ABJM theory, the theory we consider have the largest ${\cal N}=6$ supersymmetry; the same amount as the undeformed ABJM theory.
In this sence our setup is the ``simplest'' example of the non-conformally deformed field theories.
%As is the case of the four dimensional ${\cal N}=4$ supersymmetric Yang-Mills theory \cite{} or the case of the ABJM theory \cite{HMMO,CM}, a theory with (almost) maximal symmetry often enjoys unexpectedly powerful solvable structures.
Hopefully, the mass deformed ABJM theory we consider would play important roles of the toy model to reveal fundamental structures in the theories with non-trivial RG flow.

This paper is organized as follows.
In the next section we shall briefly review the mass deformation of the ABJM theory.
In section \ref{localization} we display the matrix model expression of the partition function resulting from the localization, and the saddle point equation for this matrix model.
If we formally take the mass parameter pure imaginary, the partition function coincides with that of the ABJM theory with the non-canonical $R$-charge assignments which have been studied intensively \cite{J,JKPS} in the context of the F-theorem \cite{J,CDFKS}.
To compare with our main results, we also provide the solution to the saddle point equation for the imaginary mass.
In section \ref{realmass} we solve the saddle point equation for real mass parameter.
We find two distinctive solutions and evaluate the partition function for these solutions.
In section \ref{discuss} we summarize our results with discussion and future directions.

\section{Brief review of mass deformed ABJM theory}

In this section we review the ABJM theory and its mass deformation \cite{GRVV} concerned in this paper.
The ABJM theory is the $(2+1)$ dimensional ${\cal N}=6$ $\text{U}(N)\times \text{U}(N)$ superconformal Chern-Simons theory with the Chern-Simons levels $\pm k$ coupled with four bifundamental matter fields.
In terms of the ${\cal N}=2$ superfields \cite{BKKS}, the field content of the ABJM theory consists of the two vector multiplets
\begin{align}
{\cal V}=(A_\mu,\sigma,\chi,D),\quad
{\widetilde{\cal V}}=({\widetilde A}_\mu,{\widetilde\sigma},{\widetilde\chi},{\widetilde D})
\end{align}
and four chiral multiplets $(\alpha=1,2,\dot{\alpha}=1,2)$
\begin{align}
{\cal Z}_\alpha=(A_\alpha,\varphi_\alpha,F_\alpha),\quad
{\cal W}_{\dot\alpha}=(B_{\dot\alpha},\psi_{\dot\alpha},G_{\dot\alpha})
\end{align}
which are in the bifundamental representation $(N, \bar{N})$ and $(\bar{N},N)$ of the gauge group $\text{U}(N)\times \text{U}(N)$ respectively. 
The action of the ABJM theory is written as
\begin{align}
S_\text{ABJM}=S_\text{CS}+S_\text{mat}+S_\text{pot}
\end{align}
with
\begin{align}
S_\text{CS}&=-\frac{ik}{8\pi}\int dx^3d\theta^4\int_{0}^{1}dt\Bigl[\Tr{\cal V}\bar{D}^{a}(e^{t{\cal V}}D_{a}e^{-t{\cal V}})-\Tr{\widetilde{\cal V}}\bar{D}^{a}(e^{t{\widetilde{\cal V}}}D_{a}e^{-t{\widetilde{\cal V}}})\Bigr],\nonumber \\
S_\text{mat}&=-\int dx^3d\theta^4\Bigl(\Tr\bar{\cal Z}^\alpha e^{-{\cal V}}{\cal Z}_\alpha e^{\widetilde{\cal V}}+\Tr\bar{\cal W}^{\dot{\alpha}}e^{-{\widetilde{\cal V}}}{\cal W}_{\dot{\alpha}}e^{\cal V}\Bigr),\nonumber \\
S_\text{pot}&=\frac{2\pi}{k}\int dx^3
 d\theta^2\Bigl(\Tr\epsilon^{\alpha\beta}\epsilon^{\dot{\alpha}\dot{\beta}}{\cal Z}_\alpha {\cal W}_{\dot{\alpha}}{\cal Z}_\beta {\cal W}_{\dot{\beta}}
+\Tr\epsilon_{\alpha\beta}\epsilon_{\dot{\alpha}\dot{\beta}}\bar{\cal Z}^\alpha \bar{\cal W}^{\dot{\alpha}}\bar{\cal Z}^\beta \bar{\cal W}^{\dot{\beta}}
\Bigr).
\end{align}
The chiral multiplets ${\cal Z}_\alpha$ and ${\cal W}_{\dot\alpha}$ transforms under $\text{SU}(2)\times \text{SU}(2)$ $R$-symmetry respectively.
The symmetry actually enhances to $SO(6)_R$, hence the theory have the ${\cal N}=6$ supersymmetry.
Integrating out the auxiliary fields $D$ and ${\widetilde D}$ in the vector multiplets from the Chern-Simons term $S_\text{CS}$ and the matter kinetic term $S_\text{mat}$, we obtain the following $D$-term potential for the scalars in the matter multiplets
\begin{align}
V_D=\Tr|\sigma A_\alpha-A_\alpha {\widetilde \sigma}|^2+\Tr|B_{\dot{\alpha}}\sigma-{\widetilde\sigma}B_{\dot{\alpha}}|^2
\label{Dpot}
\end{align}
together with the constraints 
\begin{align}
\sigma=\frac{2\pi}{k}(A_\alpha A^{\dagger\alpha}-B^{\dagger\dot{\alpha}}B_{\dot{\alpha}}),\quad
{\widetilde \sigma}=\frac{2\pi}{k}(A^{\dagger\alpha} A_\alpha-B_{\dot{\alpha}}B^{\dagger\dot{\alpha}}).
\label{sigmaAB}
\end{align}
Eliminating $\sigma$ and ${\widetilde\sigma}$ we indeed obtain the sextic potential for the scalars which is essential to describe the M5-branes as a fuzzy funnel as discussed in \cite{BH}.

In this paper we shall introduce the mass deformation through the following Fayet-Iliopoulos $D$-term \cite{GRVV}
\begin{align}
S_{\text{FI}}=\frac{\zeta}{2\pi} \int dx^3 d\theta^4(\Tr{\cal V}+\Tr{\widetilde{\cal V}})=\frac{\zeta}{2\pi} \int d^3x(\Tr D+\Tr {\widetilde D})
\end{align}
with the FI parameter $\zeta\in\mathbb{R}$.
Though this deformation breaks the $SO(6)_R$ symmetry down to $\text{SU}(2)\times \text{SU}(2)\times \text{U}(1)\times \mathbb{Z}_2$, the deformed theory still have ${\cal N}=6$ supersymmetry.
In this case, the constraints \eqref{sigmaAB} are shifted by the FI parameter
\begin{align}
\sigma=\frac{2\pi}{k}(A_\alpha A^{\dagger\alpha}-B^{\dagger\dot{\alpha}}B_{\dot{\alpha}})+\frac{\zeta}{k},\quad
{\widetilde \sigma}=\frac{2\pi}{k}(A^{\dagger\alpha} A_\alpha-B_{\dot{\alpha}}B^{\dagger\dot{\alpha}})-\frac{\zeta}{k}.
\end{align}
Thus the potential \eqref{Dpot} gives the mass terms with the same mass $m=\zeta/k$ to all the four scalars.
There are also some terms including the fermions and we can confirm that the theory indeed have the ${\mathcal N}=6$ supersymmetry.\footnote{
The superpotential mass term \cite{GRVV}, on the other hand, breaks some supersymmetries.
We will concentrate on the maximally supersymmetric theory in this paper for simplicity.
}

The classical vacua of the mass deformed theory was studied in \cite{GRVV,NPR}.
These vacua are also given by the matrices representing the fuzzy three sphere \cite{T,HL}.
This is an analogue of the Myers effect in the D-brane system, and the mass deformed ABJM theory represent the M2-M5 brane system where the M2-branes brow up into spherical M5-branes.

\section{Large $N$ saddle point equations}
\label{localization}

In this section we analyze the partition function of the mass deformed ABJM theory on $S^3$ with unit radius.\footnote{
We can recover the radius by the replacement $\zeta\rightarrow\zeta\cdot r_{S^3}$ since there are no other dimensionful parameters in the theory.
}
We take the limit $N\rightarrow \infty$ while keeping the level $k$ and the mass deformation $\zeta$ finite.\footnote{
As we will see later, the large $N$ behaviors may be different for $\zeta/k < 1/4$ and $\zeta/k > 1/4$. 
In this paper we shall concentrate on the former case $\zeta/k < 1/4$, which is different from the situation considered in \cite{AZ,AR}.
}
Thus we are considering the M2-branes in eleven dimensional spacetime, with finite background flux depending on $\zeta$.

The supersymmetric gauge theories on $S^3$ were studied in \cite{KWY2,J,HHL}.
With the help of the localization technique, they showed that the partition function of our theory is given by the following matrix integral
\begin{align}
Z(N)=\prod_{i=1}^N\int d\lambda_id{\widetilde\lambda}_i
e^{-f(\lambda,{\widetilde\lambda})},
\label{ZN}
\end{align}
where
\begin{align}
f(\lambda,{\widetilde\lambda})&=
\pi ik\biggl(
\sum_{i\ge 1}\lambda_i^2
-\sum_{i\ge 1}{\widetilde\lambda}_i^2
\biggr)
-2\pi i \zeta\biggl(\sum_{i\ge 1}\lambda_i+\sum_{i\ge 1}{\widetilde\lambda}_i\biggr)\nonumber \\
&\quad -\sum_{i>j}\log\sinh^2\pi(\lambda_i-\lambda_j)
-\sum_{i>j}\log\sinh^2\pi({\widetilde\lambda}_i-{\widetilde\lambda}_j)
+\sum_{i,j\ge 1}\log\cosh^2\pi(\lambda_i-{\widetilde\lambda_j}).
\label{iff}
\end{align}
Here $\lambda_{i}$ and ${\widetilde\lambda}_i$ $(i=1,\ldots,N)$ denote the eigenvalues of $\sigma$ and ${\widetilde \sigma}$ which are constant for the saddle points in the localization computation.

For the large $N$ limit, the integrals can be evaluated 
by the saddle point approximation.
The saddle point configuration 
is the solution of the following saddle point equations
\begin{align}
0&=\frac{\partial f(\lambda,{\widetilde\lambda})}{\partial
 \lambda_i}=2\pi ik\lambda_i-2\pi i \zeta-2\pi\sum_{j\neq i}\coth\pi(\lambda_i-\lambda_j)+2\pi\sum_j\tanh\pi(\lambda_i-{\widetilde\lambda}_j),\nonumber \\
0&=\frac{\partial f(\lambda,{\widetilde\lambda})}{\partial
 {\widetilde\lambda}_i}=-2\pi ik{\widetilde\lambda}_i-2\pi i \zeta-2\pi\sum_{j\neq i}\coth\pi({\widetilde\lambda}_i-{\widetilde\lambda}_j)-2\pi\sum_j\tanh\pi(\lambda_j-{\widetilde\lambda}_i).
\label{isaddle}
\end{align}
The free energy $F=-\log Z(N)$ can be evaluated by the saddle point configuration
\begin{align}
F\approx f(\lambda,{\widetilde \lambda}) |_{{\rm saddle}}.
\label{iF}
\end{align}
Note that the saddle point configurations may be complex
although $\lambda_i$ and ${\widetilde \lambda}_i$ are real in the original integration contours in \eqref{ZN}.

Below we will find the solutions of the saddle point equations
\eqref{isaddle}.
For this purpose the symmetries of the equations are helpful, as argued in \cite{HKPT}.
In the case of the ABJM theory, i.e. $\zeta=0$, the saddle point equations \eqref{isaddle} are invariant under both of the exchanges of $\lambda_i\rightarrow \pm{\widetilde\lambda}^*_i$.
Under these exchanges the free energy transforms as $f(\lambda,{\widetilde \lambda}) \rightarrow (f(\lambda,{\widetilde \lambda}))^*$.
The solutions are always paired, except for the case $\lambda_i=\pm {\widetilde\lambda}^*_i$.
Then, it is natural to assume that the lowest free energy will be realized by such self-conjugate configurations.\footnote{
If we assume a dual gravity description, paired solutions will correspond to the semi-classical solutions which are not allowed as a lowest free energy configuration.
}
This fact was indeed confirmed in the ABJM theory. 

For a general complex deformation, these exchange symmetries are broken.
There are two special choice of $\zeta$, however, where one of the $\mathbb{Z}_2$ symmetry remains.
For a real $\zeta$, which corresponds to the mass deformation, the saddle point equations are invariant under $\lambda_i \rightarrow -{\widetilde\lambda}^*_i$.
Hence we will pose the ansatz $\lambda_i =-{\widetilde\lambda}^*_i$ to solve the saddle point equations.
The other choice is the pure imaginary $\zeta$, where the remaining symmetry is $\lambda_i \rightarrow {\widetilde\lambda}^*_i$ and we should assume $\lambda_i={\widetilde\lambda}^*_i$.

\subsection{Imaginary FI-parameter (known case)}
\label{ifi}

Before going on to the real mass deformation, we shall investigate the case with pure imaginary FI-parameter\footnote{
The reader concerning only the results for mass deformed ABJM theory may skip this subsection.
}
\begin{align}
\zeta=-i\xi,\quad \xi\in\mathbb{R}.
\end{align}
Though the matrix model is equivalent to that for the $R$-charge deformation of the ABJM theory studied in \cite{J,JKPS} (see also \cite{CDFKS}), it is useful to consider these model for a demonstration of the general ideas in the evaluation of the free energy in the large $N$ limit.
Interestingly, however, the results for mass deformed ABJM theory and its ``analytically continued'' version we consider here are substantially different in various ways, contrary to the naive expectation.

\subsubsection{Analytical solution in large $N$ limit}
\label{anal}

In the case of pure imaginary FI parameter, we can find the solution to the saddle point equations \eqref{isaddle} in the large $N$ limit by evaluating the equations up to ${\cal O}(N^0)$.
As we discuss above we set
\begin{align}
\lambda_{i}={\widetilde\lambda}_i^*.
\end{align}
Furthermore, we shall assume
\begin{align}
\lambda_{i}=N^{\alpha}x_{i}+iy_{i},\quad {\widetilde\lambda}_i=N^{\alpha}x_i-iy_i,
\end{align}
with $x_{i}$ and $y_{i}$ being of order $\mathcal{O}(N^0)$.
We have introduced a factor $N^{\alpha}$ to represent the growth of the real part of the eigenvalues, where the scaling exponent $\alpha$ will be determined later.\footnote{Although it is difficult to show there are no solutions without the assumption, we believe this assumption is valid for the lowest free energy configuration, partially based on some numerical calculations.}

In the large $N$ limit, we can define continuous functions $x(s),y(s): [0,1]\rightarrow \mathbb{R}$ to replace $x_i$ and $y_i$ as
\begin{align}
\label{iassump}
 x_{i}=x\Bigl(\frac{i}{N}\Bigr),\quad  y_{i}=y\Bigl(\frac{i}{N}\Bigr).
\end{align}
Here we have ordered the eigenvalues so that $x(s)$ is a strictly increasing function.
It is more reasonable to take the real part of the eigenvalues $x$ as the fundamental variable rather than $s$ and introduce the eigenvalue density $\rho(x)$ in the $x$-direction
\begin{align}
\rho(x)=\frac{ds}{dx}
\end{align}
which is normalized to unity
\begin{align}
\int_Idx\rho(x)=1\label{unit} 
\end{align}
so that
\begin{align}
\sum_i(\cdots)_i \rightarrow N\int_Idx\rho(x)(\cdots )(x).
\end{align}
Here $I$ is the support of the eigenvalues which we shall assume to be a single finite interval $I=[a,b]$.
In the continuum notation the saddle point equations \eqref{isaddle} become
\begin{align}
\label{icsaddle}
0=-ik(N^{\alpha}x+iy(x))+\xi+N\int_{I}dx'\rho(x')\coth\pi\bigl[(x-x')N^{\alpha}+i(y(x)-y(x'))\bigl] 
 \nonumber \\
-N\int_{I}dx'\rho(x')\tanh\pi\bigr[(x-x')N^{\alpha}+i(y(x)+y(x'))\bigl].
\end{align}
We regard the integral whose integrand is singular at $x=x^{\prime}$ as the principal value integral.
Now to solve the saddle point equation \eqref{icsaddle} means to find the functions $y(x)$ and $\rho(x)$ which satisfy \eqref{icsaddle} and the normalization \eqref{unit}. 

Now we shall consider the large $N$ expansion of the last two terms in \eqref{icsaddle} including the integration over $x^\prime$.
Since the arguments of $\coth$ and $\tanh$ are scaled by $N^\alpha$, this is achieved by approximating them by the sign of the real part of the arguments and evaluating the deviations by the integration by parts.
First, we notice the following expansion formulas
\begin{align}
&\tanh(z)=\begin{cases}
\displaystyle 1-2\sum_{n=1}^{\infty}(-1)^{n-1}e^{-2nz} & ({\rm Re}(z) \geq0) \\
\displaystyle -1+2\sum_{n=1}^{\infty}(-1)^{n-1}e^{2nz} & ({\rm Re}(z)<0)
\end{cases},\nonumber \\
&\coth(z)= \begin{cases}
\displaystyle 1+2\sum_{n=1}^{\infty}e^{-2nz}  & ({\rm Re}(z) \geq0) \\
\displaystyle -1-2\sum_{n=1}^{\infty}e^{2nz} & ({\rm Re}(z)<0)
\end{cases}.
\label{iseries2}
\end{align} 
The leading terms in \eqref{iseries2} come from the sign function approximation.
In the two integrals they are precisely canceled together
\begin{align}
&N\int_{I}dx'\rho(x')\coth\pi\bigl[(x-x')N^{\alpha}+i(y(x)-y(x'))\bigl]\nonumber \\
&\quad\quad\quad\quad
\quad\quad\quad\quad
-N\int_{I}dx'\rho(x')\tanh\pi\bigr[(x-x')N^{\alpha}+i(y(x)+y(x'))\bigl]\nonumber \\
&\sim N\int_Idx^\prime\rho(x^\prime)\sgn(x-x^\prime)-N\int_Idx^\prime\rho(x^\prime)\sgn(x-x^\prime)=0.
\label{cancel}
\end{align}
Since the real part of the arguments grows with $N^{\alpha}$, the contributions from the remaining terms $e^{-2 n z}$ in \eqref{iseries2} seem to be exponentially suppressed in large $N$ limit and do not contribute to the $1/N$ expansion.
However, the contributions of the integration near $z \sim 0$ give $1/N^{\alpha}$ corrections. 
We can evaluate these terms in \eqref{iseries2} by separating the integration interval into $x>x^{\prime}$ and $x<x^{\prime}$ and integrating by parts
\begin{align}
&N\int_{I}dx'\rho(x')\coth\pi\bigl[(x-x')N^{\alpha}+i(y(x)-y(x'))\bigl]\nonumber \\
&\quad\quad\quad\quad
\quad\quad\quad\quad
-N\int_{I}dx'\rho(x')\tanh\pi\bigr[(x-x')N^{\alpha}+i(y(x)+y(x'))\bigl]\nonumber \\
&=-2iN^{1-\alpha}\rho(x)\sum_{n=1}^{\infty}\frac{(-1)^{n-1}}{\pi n}\sin(4n\pi y(x))-N^{1-2\alpha}\dot{\rho}(x)\sum_{n=1}^{\infty}\frac{(-1)^{n-1}}{\pi^2n^2}\cos(4n\pi y(x))\nonumber \\
&\quad\quad
-N^{1-2\alpha}\dot{\rho}(x)\sum_{n=1}^{\infty}\frac{1}{\pi^2n^2}+2N^{1-2\alpha}\rho(x)\dot{y}(x)\sum_{n=1}^{\infty}\frac{(-1)^{n-1}}{\pi n}\sin(4n\pi y(x))+\mathcal{O}(N^{1-3\alpha}),
\label{evalu}
\end{align}
where we have used the following formula
\begin{align}
\label{intformula}
\int_{a}^{b}g(x)e^{Ax+iy(x)}dx=\sum_{\ell=0}^{\infty}\frac{(-1)^{\ell}}{A^{\ell+1}}\biggr[\frac{d^{\ell}}{dx^{\ell}}\bigr(g(x)e^{iy(x)}\bigl)e^{Ax}\biggl]^{b}_{a}
\end{align}
with $A$ an arbitrary constant.
In our case $A$ is proportional to $N^{\alpha}$ and this formula gives the $1/N^{\alpha}$ expansion.
Here we have kept the terms up to ${\cal O}(N^{1-2\alpha})$ since these terms will be the leading contributions.

Plugging \eqref{evalu} into the saddle point equation \eqref{icsaddle},
we finally obtain two equations from the real part and the imaginary part
\begin{align}
\begin{cases}
({\rm imaginary \ part})=0&\rightarrow\quad -kN^{\alpha}x-4N^{1-\alpha}\rho(x)y(x)=0\\
({\rm real \ part})=0&\rightarrow\quad \displaystyle ky(x)+\xi-N^{1-2\alpha}\dot{\rho}(x)\Bigl[\frac{1}{4}-4y^{2}(x)\Bigr]+4N^{1-2\alpha}\rho(x)y(x)\dot{y}(x)=0
\end{cases},
\label{sadi}
\end{align}
where dot ``$\cdot$'' is the abbreviation for the differential over $x$.
We have used the following Fourier series expansion formulas by assuming $-\frac{1}{4}\leq y(x) \leq \frac{1}{4}$:
\begin{align}
\label{ifourier}
\sum_{n=1}^{\infty}\frac{(-1)^{n-1}}{n^2}\cos(4\pi
 ny)=&\frac{\pi^2}{12}-4\pi^2y^2, \quad \quad -\frac{1}{4}\leq y\leq
 \frac{1}{4}, \\
\sum_{n=1}^{\infty}\frac{(-1)^{n-1}}{n}\sin(4\pi ny)=&2\pi y, \ \quad \quad \quad \quad \quad -\frac{1}{4}\leq y\leq \frac{1}{4}.
\end{align}
Outside the range $-\frac{1}{4}\leq y(x) \leq \frac{1}{4}$, \eqref{sadi} is no longer correct.
Although we will not consider this possibility here, the formulas can be generalized by considering the periodicity of the trigonometric functions.
In order to obtain a non-trivial solution we have to balance the scalings of the two terms in the imaginary part of \eqref{sadi}, hence we shall choose
\begin{align}
\alpha=\frac{1}{2}.
\label{cons}
\end{align}
This choice also balance the scalings of all the terms in the real part of \eqref{sadi}.

Note that the non-local saddle point equations \eqref{icsaddle} have reduced to the local differential equations \eqref{sadi}.
This is because the non-local part of the equation vanishes under the assumption $\lambda_{i}={\widetilde \lambda}^*_i$, as we have seen in \eqref{cancel}. 
The saddle point equations can be solved by
\begin{align}
y(x)=-\frac{kx}{4(4\xi x+C)}, \quad  \rho(x)=4\xi x+C,
\label{isol}
\end{align}
where $C$ is an integration constant which is determined from the normalization condition \eqref{unit} as
\begin{align}
 C=\frac{1}{b-a}-2\xi(b+a).
\end{align}
Formally the solution \eqref{isol} implies $y(x)$ diverges at $x=-\frac{C}{4\xi}$, and the density $\rho(x)$ is negative for $x<-\frac{C}{4\xi}$ which obviously contradicts to the notion of the eigenvalue density.
We assume that these points are excluded from the support $I$.

Note that although we have found the solution $(y(x),\rho(x))$ the support $I$ is still completely undetermined.
The support is determined by the extremization of the free energy, as we will see in the next section.

\subsubsection{Leading behavior of free energy}
\label{imfree}
From the solution \eqref{isol} we have obtained in the last section, we can compute the free energy in the large $N$ limit \eqref{iff}.
We will obtain the free energy as a function of the edges of the support $(a,b)$.
As in the case of the ABJM theory, we can determine the support by choosing the local minimum of the free energy under the variation of $a$ and $b$.

We shall start with the continuum limit of the free energy \eqref{iff}
\begin{align}
\label{icfree}
\nonumber
 f(\lambda,{\widetilde\lambda})=&-4N^{\frac{3}{2}}\pi k\int_{I}dxx\rho(x)y(x)-4N^{\frac{3}{2}}\pi \xi\int_{I}dx\rho(x)x\\
\nonumber &-2N^{2}{\rm Re}\int_{I}dx\int_{I}dx^{\prime}\rho(x)\rho(x^{\prime})\log\sinh N^{\frac{1}{2}}\pi\bigr[(x-x^{\prime})+i(y(x)-y(x^{\prime}))\bigl]\\
&+2N^{2}\int_{I}dx\int_{I}dx^{\prime}\rho(x)\rho(x^{\prime})\log\cosh N^{\frac{1}{2}}\pi\bigr[(x-x^{\prime})+i(y(x)+y(x^{\prime}))\bigl].
\end{align}
The last two double integrations can be evaluated in the parallel way as in the last section, with the help of the formulas obtained by integrating \eqref{iseries2}
\begin{align}
&\log\cosh(z)=\begin{cases}
\displaystyle z+\sum_{n=1}^{\infty}\frac{(-1)^{n-1}e^{-2nz}}{n}-\log2 & ({\rm Re}(z) \geq0) \\
\displaystyle -z+\sum_{n=1}^{\infty}\frac{(-1)^{n-1}e^{2nz}}{n}-\log2 & ({\rm Re}(z)<0)
\end{cases},\nonumber \\
&\log\sinh(z)= \begin{cases}
\displaystyle z-\sum_{n=1}^{\infty}\frac{e^{-2nz}}{n}-\log2  & ({\rm Re}(z) \geq0) \\
\displaystyle -z-\sum_{n=1}^{\infty}\frac{e^{2nz}}{n}-\log2 \pm i\pi  & ({\rm Re}(z)<0)
\end{cases}.
\label{iseries}
\end{align} 
The contributions to the free-energy from the first terms in \eqref{iseries} are again canceled, hence there are no terms including double integration in the free energy. 
The contribution from the second terms in \eqref{iseries} are evaluated by the integration by parts with the formula \eqref{intformula} and we obtain
\begin{align}
\label{evalu2}
2\pi N^{2-\frac{1}{2}}\int_{I}dx\Bigl[\frac{1}{4}-4y^{2}(x)\Bigr]\rho^{2}(x),
\end{align}
where we have used the formula \eqref{ifourier} to reorganize the sum over $n$.
 It is enough to keep the terms up to $N^{\frac{3}{2}}$ since the
 Chern-Simons terms and FI terms are already of ${\cal O}(N^{\frac{3}{2}})$ for $\alpha=\frac{1}{2}$.

Plugging \eqref{evalu2} into \eqref{icfree} and performing the single integrations for the solution \eqref{isol}, we finally obtain
\begin{align}
\frac{f}{\pi N^{\frac{3}{2}}}=\frac{k^2(b^3-a^3)}{6}\Bigl(1-\frac{16\xi^2}{k^2}\Bigr)+\frac{1}{2(b-a)}-2\xi(b+a)+2\xi^2(b+a)^2(b-a).
\end{align}
In order for the free energy to have a local minimum with respect to $(a,b)$, the deformation $\xi$ is required to satisfy the inequality
\begin{align}
1-\frac{16\xi^2}{k^2}>0.
\end{align}
Inside this region, the values of $a$ and $b$ can be uniquely determined as\footnote{
So far we have not considered the other non-local constraint the following by integrating the real part of the saddle point equation \eqref{icsaddle}
\begin{align}
k\int_a^b dx\rho(x)y(x)+\xi=0,
\end{align}
which should have been considered before the variation of the free energy.
The solutions $(a,b)$ \eqref{isol2} indeed satisfy this condition.
}
\begin{align}
a=-\frac{1}{\sqrt{2k}}\Bigl(1-\frac{4\xi}{k}\Bigr),\quad b=\frac{1}{\sqrt{2k}}\Bigl(1+\frac{4\xi}{k}\Bigr),
\label{isol2}
\end{align}
on which the free energy is
\begin{align}
F=\frac{\sqrt{2}}{3}\pi N^{\frac{3}{2}}k^{\frac{1}{2}}\biggr(1-\frac{16\xi^2}{k^2}\biggl).
\label{resultif}
\end{align}
Substituting these vaules of $(a,b)$ into the solution \eqref{isol} we finally obtain
\begin{align}
y(x)=-\frac{kx}{4\Bigl[4\xi x+\sqrt{\frac{k}{2}}\Bigl(1-\frac{16\xi^2}{k^2}\Bigr)\Bigr]}, \quad  \rho(x)=4\xi x+\sqrt{\frac{k}{2}}\biggr(1-\frac{16\xi^2}{k^2}\biggl).
\end{align}
Note that the solution indeed satisfies the bound $-\frac{1}{4}\le y(x)\le \frac{1}{4}$ we have assumed and the positivity of the density $\rho(x)$ on the support.

Before closing this section we shall comment on the relation to the results obtained in \cite{JKPS} where the ABJM theory was deformed by assigning the non-canonical $R$-charges $\Delta$ to the bifundamental matter fields $A_i$ and $B_i$.
Our solution \eqref{isol} and \eqref{resultif} correspond to the special case of their results (See section 5 in \cite{JKPS}) with the parameters related as $\Delta_{A_{1}}=\Delta_{A_{2}}=\frac{1}{2}+\frac{2\xi}{k}$, $\Delta_{B_{1}}=\Delta_{B_{2}}=\frac{1}{2}-\frac{2\xi}{k}$ and $\Delta_{m}=0$.
The dual gravity solution was also constructed \cite{FP} which is consistent with the field theory result.

In the next section, we will consider the case of real mass with the similar method used here.
The naive guess is that the free energy and the eigenvalue distribution in the mass deformed ABJM theory would be obtained by simply replacing $\xi \rightarrow i\zeta$ and assuming $\zeta\in\mathbb{R}$.
Such an ``analytic continuation'' of the parameter, however, is not allowed generally.\footnote{
Our calculation in the large $N$ limit breaks the holomorphy in the sense that the eigenvalue distribution is separated into the real part and the imaginary part, which are real functions.
 However, it is expected that the partition function \eqref{ZN} is holomorphic at least around $\zeta=0$ from the general argument in \cite{CDFK}.
 In fact, we will confirm that the partition function is holomorphic in the large $N$ limit when the parameter is sufficiently small in section \ref{realmass}.
}
Indeed, the behavior of the matrix model \eqref{iff} greatly depends on whether the FI parameter $\zeta$ is real or imaginary.

\section{Large $N$ behavior of mass deformed ABJM theory}
\label{realmass}

In this section, we investigate the leading behavior of the mass deformed ABJM theory by the saddle point approximation.
Though the method we will use is parallel with that in the last section, the results are substantially different.
We will see that there is a solution for a real $\zeta$ where the free energy is the ``analytic continuation'' of \eqref{resultif} while the eigenvalue distribution is completely different from \eqref{isol}.
Moreover there is another solution which gives smaller free energy for $(3-2\sqrt{2})/4\leq \zeta/k \leq 1/4$, thus there occurs a phase transition as we increase $\zeta$.
These results may reflect the nontrivial structure of the vacua of the mass deformed ABJM theory.

Below we first provide the general solutions to the saddle point equations \eqref{isaddle} with real mass $\zeta\in\mathbb{R}$ in section \ref{gensol}.
We shall assume $k>0$ and $\zeta>0$ without loss of generality.
As in the case of imaginary FI parameter the solutions contains some integration constants to be determined by the non-local constraints.
In section \ref{fereal} we determine these constants and evaluate the free energy for each solutions.

\subsection{General solutions as continuous distribution}
\label{gensol}

\begin{figure}[ht!]
\begin{center}
\includegraphics[width=10cm]{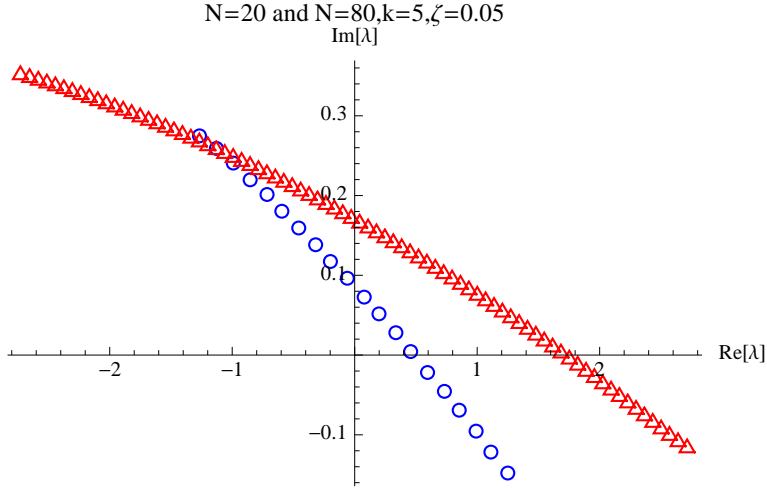}
\end{center}
\caption{
The numerical solutions $\{\lambda_i\}_{i=1}^N$ of the saddle point equation \eqref{isaddle} with reality condition \eqref{real}.
The blue circles are the eigenvalue distribution of $N=20,k=5,\zeta=0.05$, while the red triangles are that for $N=80,k=5,\zeta=0.05$.
The solutions are obtained by introducing the fictitious time $t$ and analyze late time solution $\lambda_i(t)$ of the heat equation $d\lambda_i/dt=\partial F/\partial \lambda_i$ \cite{HKPT}.
The graph indicates that the maximum of $\mathrm{Re}[\lambda_i]$ is doubled as $N$ being quadrupled.
}
\label{numerical}
\end{figure}

To solve the saddle point equations, 
we need to pose several ansatz on the eigenvalue distributions.
First we impose the following reality condition on the eigenvalues
\begin{align}
{\widetilde\lambda}_i=-\lambda_i^*,
\label{real}
\end{align}
as discussed in the previous section.
Second we switch to continuous distributions on the $x$-support $I$ as in the previous section:
\begin{align}
\label{continu}
\lambda_i\rightarrow \lambda(x)=N^{\frac{1}{2}}(x+iy(x)),\quad 
{\widetilde\lambda}_i\rightarrow {\widetilde\lambda}(x)=N^{\frac{1}{2}}(-x+iy(x)),\quad x\in I.
\end{align}
The overall scaling $N^{\frac{1}{2}}$ is observed in the numerical analysis of the saddle point equations \eqref{isaddle} (see figure \ref{numerical}).
The eigenvalue distribution in the $x$-direction is encoded in the density $\rho(x)$ normalized as \eqref{unit}.
Taking the continuum limit, the saddle point equation \eqref{isaddle} becomes
\begin{align}
0=-iN^{\frac{1}{2}}k(x+iy(x))+i\zeta+N\int_{I}dx'\rho(x')\coth\pi N^{\frac{1}{2}}\bigr[(x-x')+i(y(x)-y(x'))\bigr]\nonumber \\ 
-N\int_{I}dx'\rho(x')\tanh\pi N^{\frac{1}{2}}\bigr[(x+x')+i(y(x)-y(x'))\bigl].
\label{csaddle}
\end{align}
Now let us evaluate the last two integrations with the expansion formulas \eqref{iseries2}.
In current case the non-local contributions from the first terms in the formula \eqref{iseries2} do not cancel:
\begin{align}
N\int_{I}dx^{\prime}\rho(x^{\prime})\bigr[\sgn(x-x^{\prime})-\sgn(x+x^{\prime})\bigl].
\label{nonloc2}
\end{align}
These terms in the saddle point equation would be of ${\cal O}(N)$.
If these are no cancellation in this non-local contributions, it is impossible to solve the saddle point equations \eqref{isaddle} since there are no other terms with comparable order.
However, we can decrease the order in $N$ of the non-local contribution by assuming the support $I$ to be symmetric under the reflection $x\rightarrow -x$.
With this choice of $I$ it is reasonable to define the odd-  and the even-
function parts of $\rho(x)$ as
\begin{align}
\rho(x)=\rho_{ev}(x)+N^{-\frac{1}{2}}\rho_{od}(x),
\label{rhoevrhood}
\end{align}
where the scalings of the even/odd part are indeed required from the normalization condition \eqref{unit}
and the condition
\begin{align}
N^{\frac{1}{2}}\int_Idx\rho(x)x=\frac{\zeta}{k}
\label{imaginary}
\end{align}
which is the continuum version of the summation of the imaginary part of the saddle point equation \eqref{isaddle} over the index $i$. 
The odd part of the distribution is of ${\cal O}(N^{-\frac{1}{2}})$ because the $\text{r.h.s.}$ of \eqref{imaginary} is $N$-independent.
The non-local part in the saddle point equation \eqref{csaddle} becomes
\begin{align}
N\int_{I}dx^{\prime}\rho(x^{\prime})\bigr[\sgn(x-x^{\prime})-\sgn(x+x^{\prime})\bigl]=2N^{\frac{1}{2}}\int_{I}dx^{\prime}\rho_{od}(x)\sgn(x-x^{\prime}).
\end{align}
Though the contribution is still non-vanishing, we have achieved to reduce the order in $N$.

To solve the saddle point equations \eqref{csaddle} it is necessary to postulate the different scalings in $N$ also for the even/odd-function part of $y(x)$ as
\begin{align}
\label{continu2}
y(x)=y_{ev}(x)+N^{-\frac{1}{2}}y_{od}(x).
\end{align}
The scaling of each part is required for the consistency of the saddle point equations.
See appendix \ref{derive} for details.

Let us continue to evaluate the last two terms in \eqref{csaddle} with the formulas \eqref{iseries2}.
After substituting above ansatz, the second terms in \eqref{iseries} are evaluated by dividing the integration interval of $x^{\prime}$ into two intervals $x>x^{\prime}$ and $x<x^{\prime}$.

Then we can integrate them by parts as we
have done in section \ref{anal}. 
The leading part of the saddle point equations
\eqref{csaddle} can also be divided into four parts, the real/imaginary
and the even/odd-function parts. 
Massaging the resulting equations, we finally obtain the following four equations:
\begin{align}
0=&k\frac{d}{dx}\bigr[xy_{ev}(x)\bigl]+2\int_Idx^\prime \rho_{od}(x^\prime)\sgn(x-x^\prime),\label{saddle1}\\
0=&kx+\frac{4\rho_{ev}(y_{od}+h\sgn(x))}{1+\dot{y}_{ev}^2},\label{saddle2}\\
0=&-2k\dot{y}_{ev}(x)h{\rm sgn}(x)+\zeta(1-\dot{y}^2_{ev}(x))+\frac{1}{4}\frac{\rho_{ev}(x)\ddot{y}_{ev}(x)}{1+\dot{y}^2_{ev}(x)},\label{saddle3}\\
0=&-kh{\rm sgn}(x)-\zeta\dot{y}_{ev}(x)-\frac{1}{4}\frac{d}{dx}\biggr[\frac{\rho_{ev}(x)}{1+\dot{y}^2_{ev}(x)}\biggl]\label{saddle4},
\end{align}
where we abbreviated the differential $\frac{d}{dx}$ to dots ``$\cdot$'' as in the last section.
Here $h\in\mathbb{Z}/2$ is determined such that
\begin{align}
\label{bd}
-\frac{1}{4}\le y_{od}(x)+h\sgn(x)< \frac{1}{4}.
\end{align}
For the details of the derivation, see appendix \ref{calculations}.

Differentiating the first equation \eqref{saddle1}, we obtain a set of
four differential equations against four unknown functions $(y_{ev}(x),y_{od}(x),\rho_{ev}(x),\rho_{od}(x))$, whose general solutions are
the following two
\begin{align}
\mathrm{I}\quad&\begin{cases}
y_{ev}(x)&=-\omega|x|-\sqrt{(1+\omega^2)(x^2+a)}+b\\
y_{od}(x)&=-\frac{kx}{16\zeta\sqrt{(1+\omega^2)(x^2+a)}}-h\sgn(x)\\
\rho_{ev}(x)&=4\zeta(1+\omega^2)\Bigl[(2x^2+a)\sqrt{\frac{1+\omega^2}{x^2+a}}+2\omega|x|\Bigr]\\
\rho_{od}(x)&=\frac{kx\sqrt{1+\omega^2}(2x^2+3a)}{4(x^2+a)^{\frac{3}{2}}}+\frac{k\omega\sgn(x)}{2}
\end{cases},\nonumber \\ \nonumber \\
\mathrm{II}\quad&\begin{cases}
y_{ev}(x)&=-(\omega+\sqrt{1+\omega^2})|x|+a\\
y_{od}(x)&=-\frac{kx}{2}\frac{\sqrt{1+\omega^2}(\sqrt{1+\omega^2}+\omega)}{8\zeta|x|(1+\omega^2)(\sqrt{1+\omega^2}+\omega)+b}-h\sgn(x)\\
\rho_{ev}(x)&=8\zeta(1+\omega^2)(\sqrt{1+\omega^2}+\omega)|x|+b\\
\rho_{od}(x)&=\frac{k(\sqrt{1+\omega^2}+\omega)\sgn(x)}{2}
\end{cases},
\label{solns}
\end{align}
where
\begin{align}
\omega=\frac{hk}{\zeta}.
\end{align}
Here $a,b\in\mathbb{R}$ are integration constants.
These constants, together with the choice of the support
$I$, should be determined by the non-local constraints \eqref{unit}, \eqref{imaginary} and \eqref{saddle1}.  
We stress that there are two independent solutions.
As we shall see below, only one of these solutions is connected to that of the ABJM theory in the undeformed limit.

\subsection{Free energy}
\label{fereal}

The contribution to the free energy in the continuum limit can be written as
\begin{align}
\label{cfree}
\nonumber  f(\lambda,{\widetilde\lambda})=&-4\pi N^{2}k\int_{I}dxx\rho(x)y(x)+4N^{\frac{3}{2}}\pi\zeta\int_{I}dx\rho(x)y(x)\\
\nonumber &-2N^{2}{\rm Re}\int_{I}dx\int_{I}dx^{\prime}\rho(x)\rho(x^{\prime})\log\sinh N^{\frac{1}{2}}\pi\bigr[(x-x^{\prime})+i(y(x)-y(x^{\prime}))\bigl]\\
&+2N^{2}\int_{I}dx\int_{I}dx^{\prime}\rho(x)\rho(x^{\prime})\log\cosh N^{\frac{1}{2}}\pi\bigr[(x+x^{\prime})+i(y(x)-y(x^{\prime}))\bigl].
\end{align}
After substituting the ansatz \eqref{continu}, \eqref{rhoevrhood}  and \eqref{continu2} in section \ref{gensol}, we can evaluate the last terms in \eqref{cfree} by formula \eqref{iseries} as in section \ref{imfree} (see appendix \ref{derive} for the details):
\begin{align}
f(\lambda,{\widetilde\lambda})&=N^{\frac{3}{2}}\biggl[-4\pi k\int_{I}dx\Bigl(x\rho_{ev}(x)y_{od}(x)+x\rho_{od}(x)y_{ev}(x)\Bigr)+4\pi\zeta\int_{I}dxy_{ev}(x)\rho_{ev}(x)\nonumber \\
&\quad\quad\quad-4\pi\int_{I}dx\int_{I}dx'\rho_{od}(x)\rho_{od}(x^\prime)|x-x^\prime|\nonumber \\
&\quad\quad\quad +2\pi\int_{I}dx\frac{\rho^2_{ev}(x)}{1+\dot{y}^2_{ev}}\Bigl(\frac{1}{4}-4(y_{od}(x)+h\sgn(x))^2\Bigr)\biggr].
\label{F2}
\end{align}
In this section we compute this quantity for each solution to the saddle point equations \eqref{solns}.
For simplicity we shall assume the support $I$ to be a single segment surrounding the origin:
\begin{align}
I=(-L,L).
\label{singlesupport}
\end{align}

\subsubsection{Free energy for solution I}

To compute the free energy of the solution I in \eqref{solns}, we need
to determine the integration constants $(a,b)$ and the support $I$.
Under the assumption of single support \eqref{singlesupport}, we have
three parameters $(a,b,L)$ to be determined.
Using the three non-local constraints \eqref{unit}, \eqref{imaginary} and \eqref{saddle1} we will completely determine these parameters.\footnote{
This is in contrast to the ABJM theory and 
the $R$-charge deformation considered in section \ref{imfree} where the parameters were not completely determined from the saddle point equations but were chosen so that the free energy is minimized.
}

\begin{figure}[ht!]
\begin{center}
\includegraphics[width=10cm]{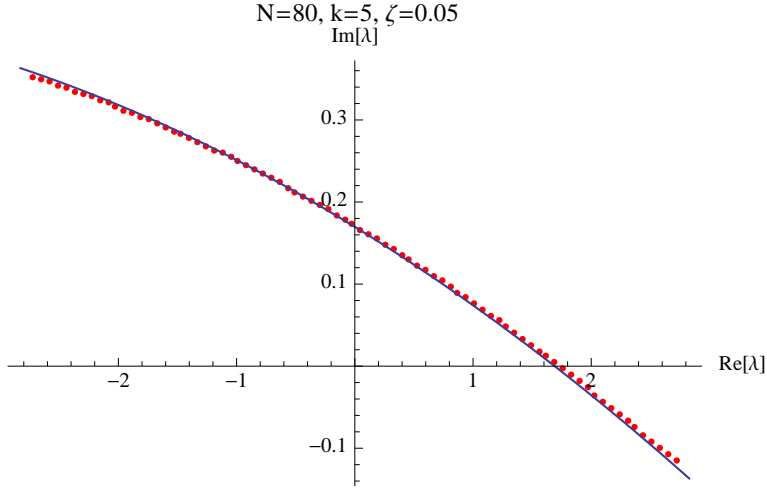}
\end{center}
\caption{
The blue line is $\lambda(x)=N^{\frac{1}{2}}(x+iy(x))$ for solution I \eqref{sol1}, while the red dots are the eigenvalue distribution obtained by a numerical analysis.
}
\label{fit}
\end{figure}

We first note that $a$ must be non-negative so that the solution is well defined on the support.
From the constraints \eqref{unit} and \eqref{imaginary} it follows that
\begin{align}
\sqrt{1+\frac{a}{L^2}}=\frac{-(X-1)\omega+\sqrt{(X-1)^2\omega^2+4X(1+\omega^2)}}{2X\sqrt{1+\omega^2}}
\end{align}
with $X=16(h^2+\zeta^2/k^2)$.
We can show that the $\text{r.h.s.}$, regarded as a function of two variables $(X,\omega)$, is always smaller than 1 for $X>1$, hence we conclude that the solution I is valid only when
\begin{align}
h=0,\quad \frac{\zeta}{k}\le \frac{1}{4}.
\label{hzeta}
\end{align}
Under these condition the three parameters are determined as
\begin{align}
a=\frac{k}{32\zeta^2}\biggl(1-\frac{16\zeta^2}{k^2}\biggr),\quad
b=\frac{\sqrt{k}}{4\sqrt{2}\zeta}\biggl(1+\frac{16\zeta^2}{k^2}\biggr),\quad
L=\frac{1}{\sqrt{2k}}.
\end{align}
After the substitution of the these values the solution I are written as
\begin{align}
\label{sol1}
y_{ev}(x)&=\frac{\sqrt{k}}{4\sqrt{2}\zeta}\biggl(1+\frac{16\zeta^2}{k^2}\biggr)-\sqrt{x^2+\frac{k}{32\zeta^2}\Bigl(1-\frac{16\zeta^2}{k^2}\Bigr)},\nonumber \\
y_{od}(x)&=- \frac{k}{16\zeta}\frac{x}{\sqrt{x^2+\frac{k}{32\zeta^2}\Bigl(1-\frac{16\zeta^2}{k^2}\Bigr)}},\nonumber \\
\rho_{ev}(x)&= 4\zeta\frac{d}{dx}\bigr(x\sqrt{x^2 + a}\bigl)=4\zeta\cdot \frac{2x^2+\frac{k}{32\zeta^2}\Bigl(1-\frac{16\zeta^2}{k^2}\Bigr)}{\sqrt{x^2+\frac{k}{32\zeta^2}\Bigl(1-\frac{16\zeta^2}{k^2}\Bigr)}},\nonumber \\ 
\rho_{od}(x)&=-\frac{k}{4}\frac{d^2}{dx^2}\bigr(xy_{ev}(x)\bigl)=\frac{kx}{4}\cdot \frac{2x^2+3\frac{k}{32\zeta^2}\Bigl(1-\frac{16\zeta^2}{k^2}\Bigr)}{\Bigl[x^2+\frac{k}{32\zeta^2}\Bigl(1-\frac{16\zeta^2}{k^2}\Bigr)\Bigr]^{\frac{3}{2}}}.
\end{align}
In figure \ref{fit} we compare the solution with the numerically
obtained eigenvalue distribution.
We can see that the numerical one coincides with the analytical one with good accuracy.

Now that the solution is completely determined, we can compute the free energy \eqref{F2} and obtain
\begin{align}
\label{result}
f_{\text{I}}=\frac{\pi\sqrt{2k}N^{\frac{3}{2}}}{3}\biggl(1+\frac{16\zeta^2}{k^2}\biggr).
\end{align}
Note that this free energy is obtained from \eqref{resultif} just by changing parameter 
$\xi \rightarrow i\zeta$ while the solution $y(x)$ and $\rho(x)$ is greatly
different from \eqref{isol} and \eqref{isol2}.\footnote{We can check that these solution are related by 
changing parameter $\xi \rightarrow i\zeta$ when we rewrite these solutions in terms of $s$ instead of $x$ (See general solutions \cite{NST2}).

}
This solution can be regarded
as the solution connected to that of the ABJM theory in the sense that the saddle point configuration and the free energy are equal to those obtained in \cite{HKPT} when we take the undeformed limit $\zeta \rightarrow 0$.
As $\zeta$ increases the free energy monotonically increases until $\zeta=\frac{k}{4}$, in contrast with \eqref{resultif}. 
However, as we will see later, 
the free energy corresponding to the solution II becomes smaller than that of the solution I as $\zeta$ crosses a certain threshold in $0<\zeta<\frac{k}{4}$.

\subsubsection{Free energy for solution II}
\begin{figure}[ht!]
\centering
\includegraphics[width=10cm]{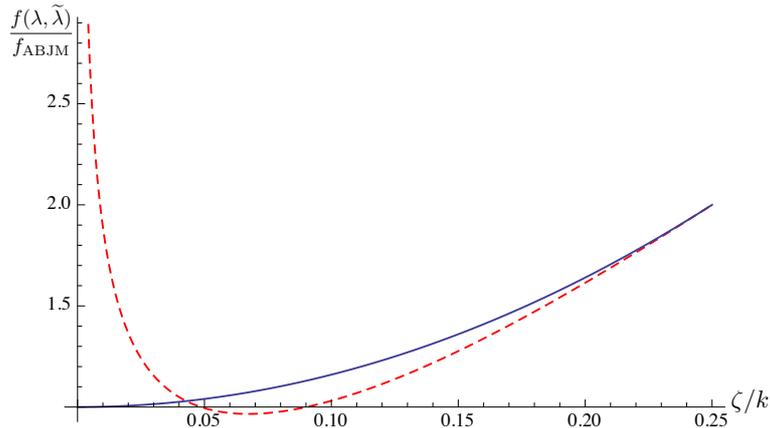}
\caption{The solid blue line is the free energy for solution I, while the dashed red line for solution II.
$f_\text{ABJM}=\pi\sqrt{2k}/3$ is the value for the ABJM theory.
The intersection point is at $\zeta/k=(3-2\sqrt{2})/4$ and at $\zeta/k=1/4$.
}
\label{fcompare}
\end{figure}

The free energy for the solution II \footnote{
This solution does not satisfy the boudary condition, which is considered as a part of the saddle point equation \cite{NST2}. The boundary condition has been ignored in previous studies.
} 
in \eqref{solns} with the assumption of single support \eqref{singlesupport} can be evaluated in similar way.
Interestingly, the same condition for $h$ and $\zeta$ \eqref{hzeta} follows from \eqref{unit}, \eqref{imaginary} and the constraint $b>0$ which is required for the positivity of the eigenvalue density $\rho_{ev}(x)$.
Together with the remaining equation \eqref{saddle1}, we can determine the three parameters $(a,b,L)$ as
\begin{align}
a=\frac{2\sqrt{2\zeta}}{k},\quad b=\frac{k}{2\sqrt{2\zeta}}\biggl(1-\frac{16\zeta^2}{k^2}\biggr),\quad L=\frac{\sqrt{2\zeta}}{k}.
\end{align}
\label{result2}
With these relations the complete expression of the solution II is
\begin{align}
y_{ev}(x)&=-|x|+\frac{2\sqrt{2\zeta}}{k},\nonumber \\
y_{od}(x)&=-\frac{kx}{16\zeta|x|+\frac{k}{\sqrt{2\zeta}}\Bigl(1-\frac{16\zeta^2}{k^2}\Bigr)},\nonumber \\ 
\rho_{ev}(x)&=8\zeta |x|+\frac{k}{2\sqrt{2\zeta}}\biggl(1-\frac{16\zeta^2}{k^2}\biggr),\nonumber \\
\rho_{od}(x)&=\frac{k}{2}\sgn(x).
\end{align}
The free energy is computed as
\begin{align}
f_{\text{II}}=\frac{\pi\sqrt{2k}N^{\frac{3}{2}}}{3}\sqrt{\frac{k}{\zeta}}\biggl(\frac{3}{16}+\frac{14\zeta^2}{k^2}-\frac{16\zeta^4}{k^4}\biggr).
\end{align}
This solution II is not connected to that of the ABJM theory since the free energy becomes infinite as $\zeta \rightarrow 0$. The free energies for the two solutions are plotted in figure \ref{fcompare}.
In the point of view of the saddle point approximation, the smallest free energy is dominant in the large $N$ limit.
In this sense $f_{I}$ is preferred when $0\le \zeta/k \leq (3-2\sqrt{2})/4$ while $f_{\text{II}}$ is preferred for $(3-2\sqrt{2})/4\leq \zeta/k \leq 1/4$. 
Therefore, we conclude that
there is a first order phase transition in the large $N$ limit of the mass deformed ABJM theory on $S^3$, with respect to the mass parameter $\zeta/k$ in this region.\footnote{
Note that the phase transition occurs only in the large $N$ limit.
For finite $N$ and finite volume, the free energy is expected to be an analytic function of $\zeta$.
}

It is interesting to consider the gravity dual of this theory.
For the imaginary FI parameter $\zeta$ or the $R$-charge deformation, the corresponding gravity 
solution in the four dimensional supergravity
was obtained in \cite{FP}
and the free energy \eqref{resultif} was reproduced.
We can see that the parameter of the dual geometry corresponding to the $R$-charge can be consistently replaced into pure imaginary, which realize the same free energy as the solution I.
Our result indicates that there exist another gravity solution corresponding to the solution II and that the phase transition also occurs in the gravity side.
We hope to investigate these points in future.

\section{Discussion}
\label{discuss}

In this paper we have calculated the large $N$ behavior of the free energy of the mass deformed ABJM theory.
With the localization method in \cite{KWY2,J,HHL}, the theory on $S^{3}$ reduces to the matrix model. 
To investigate the large $N$ behavior of the free energy we have used the
saddle point approximation to the matrix model and have solved the
saddle point equations.

The crucial point in the analysis is that we can not take the reality condition
$\lambda={\widetilde \lambda}^{*}$ in contrast to the ABJM case and the case of the $R$-charge deformation in \cite{JKPS}. 
As a result, we can not eliminate the non-local terms in the saddle point equations coming from the one-loop determinant, whatever ansatz we choose for the eigenvalue density $\rho(x)$. 
We have to consider the support of $\rho(x)$ to be symmetric: $I=[-b,b]$ to solve the saddle point equations.
Once we take $I=[-b,b]$, we can guess that it is necessary to impose the even and the odd parts of $\rho(x)$ and $y(x)$ to have different scalings in $N$.

It is also important that there are two solutions of the saddle point equations: one is connected to that of the ABJM theory while the other is not.
It depends on the value of $\zeta/k$ which solution is dominant among those two.
This is a novel phenomenon, which was absent in the ABJM theory and the theory deformed by an imaginary FI parameter where the saddle point configuratoin was uniquely determined.
Thus it would be related to the non-trivial vacuum structure of the mass deformed ABJM theory. 
We also stress that the free energy we obtained in this paper scales as
$N^{\frac{3}{2}}$ in the large $N$ limit even though the non-local contributions survive in the free energy.

There remains several problems to be concerned in future works.
In this paper we have assumed that the support $I$ of the eigenvalues is a single segment for simplicity.
It is interesting whether we can determine the solutions supported by multiple segments, and whether these solutions can be more preferable than currently obtained solutions or not.
It would also be important to reveal what occurs in the regime  $\zeta>\frac{k}{4}$ outside the bound \eqref{hzeta}.
For example, the decompactification limit of the three sphere, studied in \cite{LM}, corresponds to the limit $\zeta\rightarrow\infty$.
Since the bound automatically follows from our ansatz \eqref{continu2}, it is necessary to seek another ansatz to solve the saddle point equations.

Furthermore, there are many interesting extension of our analysis.\footnote{
See \cite{MPP} for recent analysis in gravity side.
}
For examples, the 't Hooft limit of the mass deformed ABJM theory and the theory with a boundary defined in \cite{SuTe} are interesting to be studied.
Also, though we have only considered the strict large $N$ limit, it is interesting to study the large $N$ expansion including the sub-leading corrections in $1/N$ (see recent work \cite{DF} and \cite{No} for $R$-charge deformations).
We hope to investigate these issues in near future.

\section*{Acknowledgement}
We thank to Masazumi Honda for valuable discussion.
$\text{T.N.}$ is grateful to Takaya Miyamoto for helpful comments on the numerical analysis, and also to Louise Anderson and Sanefumi Moriyama for many pieces of advice.
The work of $\text{T.N.}$ is partly supported by the JSPS Research Fellowships for Young Scientists.
Lastly, the authors would like to thank the Yukawa Institute for Theoretical Physics at Kyoto University. Discussions during the YITP workshop YITP-W-15-12 on "Development in String theory and Quantum Field Theory" were useful to complete this work.

\appendix

\section{Leading behavior of free energy in large $N$ limit}
\label{derive}
In this section, we derive the leading behavior of the free energy for real mass \eqref{F2} from \eqref{iff} in the continuum limit. We begin with the large $N$ expansion of $\lambda(x)$ and $\rho(x)$ \eqref{continu}, \eqref{continu2} with more general ansatz on Im$(\lambda)$
\begin{align}
y(x)=N^{\alpha}y_{ev}(x)+N^{\beta}y_{od}(x),\quad
\end{align}
which reduces to the ansatz \eqref{continu} when
\begin{align}
\label{const}
\alpha=0,\quad \beta=-\frac{1}{2}.
\end{align}
Below we argue that the exponent $(\alpha,\beta)$ are fixed to these values if we require that the contribution of the free energy from the Chern-Simons terms, FI terms and the one-loop determinant terms to balance each other.
With these conditions, we also find that the free energy indeed scales as $f(\lambda,{\widetilde\lambda}) \sim
N^{\frac{3}{2}}$. We further collect all the leading terms in the large
$N$ limit to decide leading coefficient of the free energy as
\eqref{F2}.

In the continuum limit, the free energy \eqref{iff} becomes
\begin{align}
\label{cfreea}
\nonumber  f(\lambda,{\widetilde\lambda})=&-4\pi k\int_{I}dx\bigr[N^{2+\beta}x\rho_{ev}(x)y_{od}(x)+N^{\frac{3}{2}+\alpha}x\rho_{od}(x)y_{ev}(x)\bigl]\\
\nonumber &+4N^{\frac{3}{2}+\alpha}\pi \zeta\int_{I}dx\rho_{ev}(x)y_{ev}(x)\\
\nonumber &-2N^{2}{\rm Re}\int_{I}dx\int_{I}dx^{\prime}\rho(x)\rho(x^{\prime})\log\sinh N^{\frac{1}{2}}\pi\bigr[(x-x^{\prime})+i(y(x)-y(x^{\prime}))\bigl]\\
&+2N^{2}\int_{I}dx\int_{I}dx^{\prime}\rho(x)\rho(x^{\prime})\log\cosh N^{\frac{1}{2}}\pi\bigr[(x+x^{\prime})+i(y(x)-y(x^{\prime}))\bigl].
\end{align}
The last two terms containing double integration can be evaluated by applying the expansion formulas \eqref{iseries}.
The contributions from the first terms in \eqref{iseries} can be explicitly evaluated and found to scale as $N^\frac{3}{2}$
\begin{align}
\label{nonlocal}
&-2N^{\frac{5}{2}}\pi\int_{I}dx\int_{I}dx^{\prime}\rho(x)\rho(x^{\prime})|x-x^{\prime}|+2N^{\frac{5}{2}}\pi\int_{I}dx\int_{I}dx^{\prime}\rho(x)\rho(x^{\prime})|x+x^{\prime}|\nonumber \\
&=-4N^{\frac{3}{2}}\pi\int_{I}dx\int_{I}dx^{\prime}\rho_{od}(x)\rho_{od}(x^{\prime})|x-x^{\prime}|,
\end{align}
where we have neglected the imaginary part since the reality of the  free energy is manifest.
On the other hand, the contributions coming from the second terms in \eqref{iseries} can be estimated as $N^{2-\frac{1}{2}}=N^{\frac{3}{2}}$ since the exponents are of order ${\cal O}(N^{\frac{1}{2}})$. Therefore the balance condition is
\begin{align}
N^{\frac{3}{2}+\alpha}=N^{2+\beta}=N^{\frac{3}{2}},
\end{align}
which determine $\alpha$ and $\beta$ as \eqref{const}.

Now we look the contributions from the second terms in \eqref{iseries} in more detail
\begin{align}
\label{delta}
&2N^{2}{\rm Re}\int_{I}dx\int_{I}dx^{\prime}\rho(x)\rho(x^{\prime})\sum_{n=1}^{\infty}\frac{1}{n}e^{-2nN^{\frac{1}{2}}\pi \sgn(x-x^{\prime})(x-x^{\prime}+i(y(x)-y(x^{\prime})))}\nonumber \\
&\quad\quad\quad +2N^{2}\int_{I}dx\int_{I}dx^{\prime}\rho(x)\rho(x^{\prime})\sum_{n=1}^{\infty}\frac{(-1)^{n-1}}{n}e^{-2nN^{\frac{1}{2}}\pi \sgn(x+x^{\prime})(x+x^{\prime}+i(y(x)-y(x^{\prime})))}.
\end{align}
Separating the integration domain of $x^{\prime}$ into $x>x^{\prime}$ and $x<x^{\prime}$ and integrating by parts, we obtain 
\begin{align}
2N^{\frac{3}{2}}\int_{I}dx\frac{\rho^2_{ev}(x)}{1+\dot{y}^2_{ev}(x)}\sum_{n=1}^{\infty}\biggr[\frac{(-1)^{n-1}}{\pi n^2}\cos(4n\pi y_{od}(x))+\frac{1}{\pi n^2}\biggr]+\mathcal{O}(N^{1}).
\end{align}
Here dot ``$\cdot$'' stands for the differential $\partial_x$.
In the integration by parts, we have used the following formula:
\begin{align}
\label{formula}
\int_{a}^{b}g(x)e^{A(x+iy(x))}dx=&\sum_{\ell=0}^{\infty}\frac{1}{A^{\ell+1}}\biggr[\frac{1}{1+i
 \dot{y}(x)}(D^\ell g(x))e^{A(x+iy(x))}\biggl]^{b}_{a}, \quad
Dg(x) \equiv -\frac{d}{dx}\biggl(\frac{g(x)}{1+i\dot{y}(x)}\biggr),
\end{align}
where $A$ is some constant, which follows from the Leibniz rule 
\begin{align}
g(x)e^{A(x+iy(x))}=\frac{d}{dx}\biggr[\frac{g(x)e^{A(x+iy(x))}}{A(1+i \dot{y}(x)\bigl)}\biggl]+\frac{1}{A}\bigr(Dg(x)\bigl)e^{A(x+iy(x))}.
\end{align}
In this case the constant $A$ is proportional to $N^{\frac{1}{2}}$, thus the formula \eqref{formula} implies that the integrals can be approximated as the sign functions and have $N^{-\frac{1}{2}}$ correction.
In the derivation of \eqref{delta}, the boundary terms from $x^\prime=\pm L$ are ignored since they are exponentially suppressed ${\cal O}(e^{-N^{\frac{1}{2}}L})$ compared with the boundary terms from $x^\prime=\pm x$.
We can compute the sum over $n$ with the help of the formula
\begin{align}
\label{fourier}
\sum_{n=1}^{\infty}\frac{(-1)^{n-1}}{n^2}\cos(4\pi ny)=\frac{\pi^2}{12}-4\pi^2(y+h)^2, \quad \quad -\frac{1}{4}\leq y+h\leq \frac{1}{4},\quad h\in\mathbb{Z}/2,
\end{align}
and obtain 
\begin{align}
2N^{\frac{3}{2}}\pi\int_{I}dx\frac{\rho^2_{ev}(x)}{1+\dot{y}^2_{ev}(x)}\Bigr[\frac{1}{4}-(y_{od}(x)+h\sgn(x))^2\Bigl].
\end{align}
Here the shift parameter $h$ is introduced as in \eqref{bd}. Plugging this result and the result \eqref{nonlocal} into \eqref{cfreea}, we obtain the leading coefficient of the free energy \eqref{F2}

\section{Evaluation of saddle point equation in large $N$ limit}
\label{calculations}
In this section, we derive the  saddle point equations \eqref{saddle1}, \eqref{saddle2}, \eqref{saddle3} and \eqref{saddle4} in detail with the assumptions \eqref{continu}, \eqref{rhoevrhood} and \eqref{continu2}.
The computations in this section are parallel with those in appendix \ref{derive}.
In the continuum limit, the saddle point equation \eqref{isaddle} becomes 
\begin{align}
0=-iN^{\frac{1}{2}}k(x+iy(x))+i\zeta+N\int_{I}dx'\rho(x')\coth\pi N^{\frac{1}{2}}\bigr[(x-x')+i(y(x)-y(x'))\bigr]\nonumber \\ 
-N\int_{I}dx'\rho(x')\tanh\pi N^{\frac{1}{2}}\bigr[(x+x')+i(y(x)-y(x'))\bigl].\label{csaddlea}
\end{align}

We now evaluate the last two integrations by using the expansion formulas \eqref{iseries2}.
The contributions to the saddle point equation from the first terms in these formulas are of order $N^{1-\frac{1}{2}}$:
\begin{align}
\label{nonloc2}
N\int_{I}dx^{\prime}\rho(x^{\prime})\bigr[\sgn(x-x^{\prime})-\sgn(x+x^{\prime})\bigl]=2N^{\frac{1}{2}}\int_{I}dx^{\prime}\rho_{od}(x^{\prime})\sgn(x-x^{\prime}).
\end{align}

The deviations from the sign functions can be computed by using the formula \eqref{formula}, which are $N^{-\frac{1}{2}}$ corrections.
The saddle point equation is decomposed to the real and the imaginary parts.
Moreover, since we separated $\rho(x)$ and $y(x)$ into the even and the
odd-function parts, the saddle point equation is also separated to the
even and the odd parts.
It can be seen that the leading contributions of 
these four separated parts have different scalings in $N$.
To obtain all of the four leading parts, 
we need to evaluate the deviations up to ${\cal O}(N^0)$:

\begin{align}
&2N\int_{I}dx^{\prime}\rho(x^{\prime})\sgn (x-x^{\prime})\sum_{n=1}^{\infty}e^{-2n\pi N^{\frac{1}{2}}\sgn(x-x^{\prime})\bigr[x-x^{\prime}+i(y(x)-y(x^{\prime}))\bigl]}\nonumber \\
&\quad\quad\quad\quad\quad\quad\quad\quad+2N\int_{I}dx^{\prime}\rho(x^{\prime})\sgn(x+x^{\prime})\sum_{n=1}^{\infty}(-1)^{n-1}e^{-2n\pi N^{\frac{1}{2}}\sgn(x+x^{\prime})\bigr[x+x^{\prime}+i(y(x)-y(x^{\prime}))\bigl]}\\
\nonumber =&-N^{\frac{1}{2}}\sum_{n=1}^{\infty}\frac{2i\rho(-x)}{1-i\dot{y}(-x)}\frac{(-1)^{n-1}}{n\pi}\sin(4n\pi y_{od}(x))\\
&\nonumber \quad\quad\quad+\sum_{n=1}^{\infty}\frac{1}{\pi^{2}}\frac{1}{1-i\dot{y}(-x)}\frac{d}{dx^{\prime}}\biggl[\frac{ \rho(x^\prime)}{1-i\dot{y}(x^{\prime})}\biggr]_{x'=-x}\biggr[\frac{(-1)^{n-1}}{n^{2}}\cos(4n\pi y_{od}(x))\biggl]\nonumber \\
&\quad\quad\quad-\sum_{n=1}^{\infty}\frac{1}{\pi^{2}n^2}\frac{1}{1+i\dot{y}(x)}\frac{d}{dx^{\prime}}\biggl[\frac{\rho(x^\prime)}{1+i\dot{y}(x^{\prime})}\biggr]_{x'=x}.
\end{align}
Substituting this and \eqref{nonloc2}, we obtain the leading
contributions of saddle point equation in which 
the sum over $n$ can be reorganized 
with the formula \eqref{fourier}.
Finally we find that the leading terms in the
saddle point equation \eqref{csaddlea} decompose as
\begin{align}
0=N^{\frac{1}{2}}G_{r,e}+G_{r,o}+i\bigr(N^{\frac{1}{2}}G_{i,o}+G_{i,e}\bigl)+\hdots
\end{align}
where $G_{r,e}$($G_{i,e}$) and $G_{r,o}$($G_{i,o}$) denote the leading contributions of even and odd part of the real (imaginary) part of the saddle point equation and are defined as
\begin{align}
G_{r,e}&=ky_{ev}+2\int_Idx^\prime\rho_{od}(x^\prime)\sgn(x-x^\prime)-\frac{4\rho_{ev}(y_{od}+h\sgn(x))\dot{y}_{ev}}{1+\dot{y_{ev}^2}},\label{difeq1} \\
G_{r,o}&=ky_{od}
+\frac{4(\rho_{ev}\dot{y}_{od}+\rho_{od}\dot{y}_{ev})(y_{od}+h\sgn(x))}
{1+\dot{y}_{ev}^2}
-\frac{8\rho_{ev}(y_{od}+h\sgn(x))\dot{y}_{ev}^2\dot{y}_{od}}
{(1+\dot{y}_{ev}^2)^2}\nonumber \\
&\quad\quad -\biggl[\frac{\dot{\rho}_{ev}(1-\dot{y}_{ev}^2)}{(1+\dot{y}_{ev}^2)^2}-\frac{\rho_{ev}\ddot{y}_{ev}(3\dot{y}_{ev}-\dot{y}_{ev}^3)}{(1+\dot{y}_{ev}^2)^3}\biggr](\frac{1}{4}-4(y_{od}+h{\rm sgn}(x))^2),\label{difeq2}\\
G_{i,e}&=\zeta+\frac{4\rho_{od}(y_{od}+h\sgn(x))}{1+\dot{y}_{ev}^2}
-\frac{8\rho_{ev}(y_{od}+h\sgn(x))\dot{y}_{ev}\dot{y}_{od}}{(1+\dot{y}_{ev}^2)^2}\nonumber \\
&\quad\quad\quad\quad\quad\quad-\biggl[-4(y_{od}+h\sgn(x))^2-\frac{1}{4}\biggr]\biggl[
-\frac{2\dot{\rho}_{ev}\dot{y}_{ev}}{(1+\dot{y}_{ev}^2)^2}
+\frac{\rho_{ev}\ddot{y}_{ev}(-1+3\dot{y}_{ev}^2)}{(1+\dot{y}_{ev}^2)^3}
\biggr],\label{difeq3} \\
G_{i,o}&=-kx-\frac{4\rho_{ev}(y_{od}+h\sgn(x))}{1+\dot{y}_{ev}^2}.
\label{difeq4}
\end{align}

These equations can be simplified to 
the set of the equations \eqref{saddle1}, \eqref{saddle2}, \eqref{saddle3}
and \eqref{saddle4}, as follows.
The first equation \eqref{saddle1} can be obtained by considering
\begin{align}
0=G_{r,e}-\dot{y}_{ev}G_{i,o}.
\end{align}
To derive the third and fourth equations, \eqref{saddle3} and
\eqref{saddle4}, we notice the following relation obtained by differentiating \eqref{saddle1}
\begin{align}
0=k\frac{d^2}{dx^2}\bigr[xy_{ev}(x)\bigl]+4\rho_{od}(x),
\label{dsaddle1}
\end{align}
and consider the following two combinations
\begin{align}
0&=2\dot{y}_{ev}(x)G_{r,o}+(1-\dot{y}^2_{ev}(x))G_{i,e},\nonumber \\
0&=G_{r,o}-\dot{y}_{ev}G_{i,e}.
\end{align}
The explicit form of the first combination is
\begin{align}
2\dot{y}_{ev}(x)G_{r,o}+(1-\dot{y}^2_{ev}(x))G_{i,e}&=
2k\dot{y}_{ev}(x)y_{od}(x)+\zeta (1-\dot{y}^2_{ev}(x))+4\rho_{od}(x)\bigr(y_{od}(x)+h{\rm sgn}(x)\bigl)\nonumber \\ 
&\!\!\!\!\!\!\!\!\!\!\!\!+\frac{1}{4}\frac{\rho_{ev}(x)\ddot{y}_{ev}}{1+\dot{y}^2_{ev}(x)}-4\frac{\rho_{ev}(x)\bigr(y_{od}(x)+h{\rm sgn}(x)\bigl)}{1+\dot{y}^2_{ev}(x)}\ddot{y}_{ev}(x)\bigr(y_{od}(x)+h{\rm sgn}(x)\bigl)
\end{align}
which reduces to the $\text{r.h.s.}$ of \eqref{saddle3} with the help of \eqref{difeq4} and \eqref{dsaddle1}, while the second combination is
\begin{align}
&G_{r,o}-\dot{y}_{ev}G_{i,e}\nonumber \\
&=ky_{od}(x)-\zeta \dot{y}_{ev}(x)-\frac{1}{4}\frac{d}{dx}\biggr[\frac{\rho_{ev}(x)}{1+\dot{y}^2_{ev}(x)}\biggl]+4\frac{d}{dx}\biggr[\frac{{\rho}_{ev}(x)(y_{od}(x)+h{\rm sgn}(x))}{1+\dot{y}^2_{ev}(x)}\biggl](y_{od}(x)+h{\rm sgn}(x))
\end{align}
which reduces to the $\text{r.h.s.}$ of \eqref{saddle4} due to \eqref{difeq4}.


\begin{thebibliography}{99}
%\cite{Aharony:2008ug}
\bibitem{ABJM} 
  O.~Aharony, O.~Bergman, D.~L.~Jafferis and J.~Maldacena,
  ``N=6 superconformal Chern-Simons-matter theories, M2-branes and their gravity duals,''
  JHEP {\bf 0810}, 091 (2008)
  [arXiv:0806.1218 [hep-th]].
  %%CITATION = ARXIV:0806.1218;%%
%
\bibitem{T} 
  S.~Terashima,
  ``On M5-branes in N=6 Membrane Action,''
  JHEP {\bf 0808}, 080 (2008)
  [arXiv:0807.0197 [hep-th]].
  %%CITATION = ARXIV:0807.0197;%%
%
\bibitem{NT}
  T.~Nosaka and S.~Terashima,
  ``M5-branes in ABJM theory and Nahm equation,''
  Phys.\ Rev.\ D {\bf 86} (2012) 125027
  doi:10.1103/PhysRevD.86.125027
  [arXiv:1208.1108 [hep-th]].
  %%CITATION = doi:10.1103/PhysRevD.86.125027;%%
%
\bibitem{ST}
  K.~Sakai and S.~Terashima,
  ``Integrability of BPS equations in ABJM theory,''
  JHEP {\bf 1311} (2013) 002
  doi:10.1007/JHEP11(2013)002
  [arXiv:1308.3583 [hep-th]].
  %%CITATION = doi:10.1007/JHEP11(2013)002;%%
%
\bibitem{BH}
  A.~Basu and J.~A.~Harvey,
  ``The M2-M5 brane system and a generalized Nahm's equation,''
  Nucl.\ Phys.\ B {\bf 713} (2005) 136
  doi:10.1016/j.nuclphysb.2005.02.007
  [hep-th/0412310].
  %%CITATION = doi:10.1016/j.nuclphysb.2005.02.007;%%
%
\bibitem{BL1}
  J.~Bagger and N.~Lambert,
  ``Modeling Multiple M2's,''
  Phys.\ Rev.\ D {\bf 75} (2007) 045020
  doi:10.1103/PhysRevD.75.045020
  [hep-th/0611108].
  %%CITATION = doi:10.1103/PhysRevD.75.045020;%%
%
\bibitem{G}
  A.~Gustavsson,
  ``Algebraic structures on parallel M2-branes,''
  Nucl.\ Phys.\ B {\bf 811} (2009) 66
  doi:10.1016/j.nuclphysb.2008.11.014
  [arXiv:0709.1260 [hep-th]].
  %%CITATION = doi:10.1016/j.nuclphysb.2008.11.014;%%
%
\bibitem{BL2}
  J.~Bagger and N.~Lambert,
  ``Gauge symmetry and supersymmetry of multiple M2-branes,''
  Phys.\ Rev.\ D {\bf 77} (2008) 065008
  doi:10.1103/PhysRevD.77.065008
  [arXiv:0711.0955 [hep-th]].
  %%CITATION = doi:10.1103/PhysRevD.77.065008;%%
%
\bibitem{BL3}
  J.~Bagger and N.~Lambert,
  ``Comments on multiple M2-branes,''
  JHEP {\bf 0802} (2008) 105
  doi:10.1088/1126-6708/2008/02/105
  [arXiv:0712.3738 [hep-th]].
  %%CITATION = doi:10.1088/1126-6708/2008/02/105;%%
%
\bibitem{GRVV} 
  J.~Gomis, D.~Rodriguez-Gomez, M.~Van Raamsdonk and H.~Verlinde,
  ``A Massive Study of M2-brane Proposals,''
  JHEP {\bf 0809}, 113 (2008)
  [arXiv:0807.1074 [hep-th]].
  %%CITATION = ARXIV:0807.1074;%%
%
\bibitem{HL} 
  K.~Hanaki and H.~Lin,
  ``M2-M5 Systems in N=6 Chern-Simons Theory,''
  JHEP {\bf 0809}, 067 (2008)
  [arXiv:0807.2074 [hep-th]].
  %%CITATION = ARXIV:0807.2074;%%
%
\bibitem{TY1}
  S.~Terashima and F.~Yagi,
  ``M5-brane Solution in ABJM Theory and Three-algebra,''
  JHEP {\bf 0912} (2009) 059
  doi:10.1088/1126-6708/2009/12/059
  [arXiv:0909.3101 [hep-th]].
  %%CITATION = doi:10.1088/1126-6708/2009/12/059;%%
%
\bibitem{TY2}
  S.~Terashima and F.~Yagi,
  ``On Effective Action of Multiple M5-branes and ABJM Action,''
  JHEP {\bf 1103} (2011) 036
  doi:10.1007/JHEP03(2011)036
  [arXiv:1012.3961 [hep-th]].
  %%CITATION = doi:10.1007/JHEP03(2011)036;%%
%
\bibitem{NPR} 
  H.~Nastase, C.~Papageorgakis and S.~Ramgoolam,
  ``The Fuzzy S**2 structure of M2-M5 systems in ABJM membrane theories,''
  JHEP {\bf 0905}, 123 (2009)
  [arXiv:0903.3966 [hep-th]].
  %%CITATION = ARXIV:0903.3966;%%
%
\bibitem{M}
  R.~C.~Myers,
  ``Dielectric branes,''
  JHEP {\bf 9912} (1999) 022
  doi:10.1088/1126-6708/1999/12/022
  [hep-th/9910053].
  %%CITATION = doi:10.1088/1126-6708/1999/12/022;%%
%
\bibitem{BW}
  I.~Bena and N.~P.~Warner,
  ``A Harmonic family of dielectric flow solutions with maximal supersymmetry,''
  JHEP {\bf 0412} (2004) 021
  doi:10.1088/1126-6708/2004/12/021
  [hep-th/0406145].
  %%CITATION = doi:10.1088/1126-6708/2004/12/021;%%
%
\bibitem{PTW}
  K.~Pilch, A.~Tyukov and N.~P.~Warner,
  ``Flowing to Higher Dimensions: A New Strongly-Coupled Phase on M2 Branes,''
  arXiv:1506.01045 [hep-th].
  %%CITATION = ARXIV:1506.01045;%%
%
\bibitem{P}
  V.~Pestun,
  ``Localization of gauge theory on a four-sphere and supersymmetric Wilson loops,''
  Commun.\ Math.\ Phys.\  {\bf 313} (2012) 71
  doi:10.1007/s00220-012-1485-0
  [arXiv:0712.2824 [hep-th]].
  %%CITATION = doi:10.1007/s00220-012-1485-0;%%
%
\bibitem{W}
  E.~Witten,
  ``Topological Quantum Field Theory,''
  Commun.\ Math.\ Phys.\  {\bf 117} (1988) 353.
  doi:10.1007/BF01223371
  %%CITATION = doi:10.1007/BF01223371;%%
%
\bibitem{N}
  N.~A.~Nekrasov,
  ``Seiberg-Witten prepotential from instanton counting,''
  Adv.\ Theor.\ Math.\ Phys.\  {\bf 7} (2003) 5,  831
  doi:10.4310/ATMP.2003.v7.n5.a4
  [hep-th/0206161].
  %%CITATION = doi:10.4310/ATMP.2003.v7.n5.a4;%%
%
\bibitem{KWY2}
  A.~Kapustin, B.~Willett and I.~Yaakov,
  ``Nonperturbative Tests of Three-Dimensional Dualities,''
  JHEP {\bf 1010} (2010) 013
  [arXiv:1003.5694 [hep-th]].
  %%CITATION = ARXIV:1003.5694;%%
%
\bibitem{J} 
  D.~L.~Jafferis,
  ``The Exact Superconformal R-Symmetry Extremizes Z,''
  JHEP {\bf 1205}, 159 (2012)
  [arXiv:1012.3210 [hep-th]].
  %%CITATION = ARXIV:1012.3210;%%
%
\bibitem{HHL} 
  N.~Hama, K.~Hosomichi and S.~Lee,
  ``Notes on SUSY Gauge Theories on Three-Sphere,''
  JHEP {\bf 1103}, 127 (2011)
  [arXiv:1012.3512 [hep-th]].
  %%CITATION = ARXIV:1012.3512;%%
%
\bibitem{JKPS}
  D.~L.~Jafferis, I.~R.~Klebanov, S.~S.~Pufu and B.~R.~Safdi,
  ``Towards the F-Theorem: N=2 Field Theories on the Three-Sphere,''
  JHEP {\bf 1106}, 102 (2011)
  [arXiv:1103.1181 [hep-th]].
  %%CITATION = ARXIV:1103.1181;%%
%
\bibitem{CDFKS}
  C.~Closset, T.~T.~Dumitrescu, G.~Festuccia, Z.~Komargodski and N.~Seiberg,
  ``Contact Terms, Unitarity, and F-Maximization in Three-Dimensional Superconformal Theories,''
  JHEP {\bf 1210} (2012) 053
  doi:10.1007/JHEP10(2012)053
  [arXiv:1205.4142 [hep-th]].
  %%CITATION = doi:10.1007/JHEP10(2012)053;%%
%
\bibitem{BKKS} 
  M.~Benna, I.~Klebanov, T.~Klose and M.~Smedback,
  ``Superconformal Chern-Simons Theories and AdS(4)/CFT(3) Correspondence,''
  JHEP {\bf 0809}, 072 (2008)
  [arXiv:0806.1519 [hep-th]].
  %%CITATION = ARXIV:0806.1519;%%
%
\bibitem{AZ} 
  L.~Anderson and K.~Zarembo,
  ``Quantum Phase Transitions in Mass-Deformed ABJM Matrix Model,''
  JHEP {\bf 1409}, 021 (2014)
  doi:10.1007/JHEP09(2014)021
  [arXiv:1406.3366 [hep-th]].
  %%CITATION = doi:10.1007/JHEP09(2014)021;%%
%
\bibitem{AR} 
  L.~Anderson and J.~G.~Russo,
  ``ABJM Theory with mass and FI deformations and Quantum Phase Transitions,''
  JHEP {\bf 1505}, 064 (2015)
  [arXiv:1502.06828 [hep-th]].
  %%CITATION = ARXIV:1502.06828;%%
%
\bibitem{HKPT}
  C.~P.~Herzog, I.~R.~Klebanov, S.~S.~Pufu and T.~Tesileanu,
  ``Multi-Matrix Models and Tri-Sasaki Einstein Spaces,''
  Phys.\ Rev.\ D {\bf 83}, 046001 (2011)
  [arXiv:1011.5487 [hep-th]].
  %%CITATION = ARXIV:1011.5487;%%
%
\bibitem{FP}
  D.~Z.~Freedman and S.~S.~Pufu,
  ``The holography of $F$-maximization,''
  JHEP {\bf 1403} (2014) 135
  doi:10.1007/JHEP03(2014)135
  [arXiv:1302.7310 [hep-th]].
  %%CITATION = doi:10.1007/JHEP03(2014)135;%%
%
\bibitem{CDFK}
  C.~Closset, T.~T.~Dumitrescu, G.~Festuccia and Z.~Komargodski,
  ``From Rigid Supersymmetry to Twisted Holomorphic Theories,''
  Phys.\ Rev.\ D {\bf 90} (2014) 8,  085006
  doi:10.1103/PhysRevD.90.085006
  [arXiv:1407.2598 [hep-th]].
  %%CITATION = doi:10.1103/PhysRevD.90.085006;%%
%
\bibitem{LM}
  H.~Lin and J.~M.~Maldacena,
  ``Fivebranes from gauge theory,''
  Phys.\ Rev.\ D {\bf 74} (2006) 084014
  doi:10.1103/PhysRevD.74.084014
  [hep-th/0509235].
  %%CITATION = doi:10.1103/PhysRevD.74.084014;%%
%
%\cite{Massai:2014wba}
\bibitem{MPP} 
  S.~Massai, G.~Pasini and A.~Puhm,
  ``Metastability in Bubbling AdS Space,''
  JHEP {\bf 1502}, 138 (2015)
  %doi:10.1007/JHEP02(2015)138
  [arXiv:1407.6007 [hep-th]].
  %%CITATION = doi:10.1007/JHEP02(2015)138;%%
  %5 citations counted in INSPIRE as of 22 May 2016
\bibitem{SuTe}
  S.~Sugishita and S.~Terashima,
  ``Exact Results in Supersymmetric Field Theories on Manifolds with Boundaries,''
  JHEP {\bf 1311} (2013) 021
  doi:10.1007/JHEP11(2013)021
  [arXiv:1308.1973 [hep-th]].
  %%CITATION = doi:10.1007/JHEP11(2013)021;%%
%
\bibitem{DF}
  N.~Drukker and J.~Felix,
  ``3d mirror symmetry as a canonical transformation,''
  JHEP {\bf 1505} (2015) 004
  doi:10.1007/JHEP05(2015)004
  [arXiv:1501.02268 [hep-th]].
  %%CITATION = doi:10.1007/JHEP05(2015)004;%%
%
\bibitem{No}
  T.~Nosaka,
  ``Instanton effects in ABJM theory with general R-charge assignments,''
  arXiv:1512.02862 [hep-th].
  %%CITATION = ARXIV:1512.02862;%%
%
\bibitem{NST2}
  T.~Nosaka, K.~Shimizu and S.~Terashima,
  ``Mass Deformed ABJM Theory on Three Sphere in Large N limit,''
  arXiv:1608.02654 [hep-th].
  %%CITATION = ARXIV:1608.02654;%%
\end{thebibliography}
\end{document}